\documentclass[english,aps,prd,showpacs,twocolumn,preprintnumbers,nofootinbib]{revtex4-1}

\usepackage{amsmath}
\usepackage{amssymb}

\usepackage{slashed}
\usepackage{graphicx}
\usepackage{epsfig}

\usepackage{epstopdf}
\usepackage{esint}

\begin{document}
\preprint{Alberta Thy 9-15}

\title{Shape function in QED and bound muon decays}
\author{Robert Szafron}
\author{Andrzej Czarnecki}
\affiliation{Department of Physics, University of Alberta, Edmonton,
  Alberta, Canada T6G 2G7}

\begin{abstract}
 When a particle decays in an external field, the energy spectrum of
 the products is smeared. We derive an analytical expression for the
 shape function accounting for the motion of the decaying particle
 and the final state interactions.  We apply our result to calculate
 the muonium decay spectrum and comment on applications to the muon
 bound in an atom.
\end{abstract}


\maketitle

\section{Introduction}

A bound particle decays differently than when it is free. Even in the
ground state, due to the uncertainty principle, bound particles are in
motion that causes a Doppler smearing of their decay
products. In addition, if the charge responsible for the binding is
conserved, daughter particles are subject to  final state
interactions.

Binding effects partially cancel in the total decay width
\cite{Uberall:1960zz,Czarnecki:1999yj,Chay:1990da,
  Bigi:1992su}. However, in some  regions, the
energy spectrum of the decay products can be significantly
deformed.  The range of the accessible energy can also be modified, by
a participation of  spectators.

In this paper we focus on weakly bound systems in quantum
electrodynamics (QED) where the bulk of the decay products remains in
the energy range accessible also in the free decay. The slight but
interesting redistribution in that region is governed by the so-called
shape function $S$ \cite{Neubert:1993ch, Neubert:1993um,  Bigi:1993ex,
  Mannel:1994pm, DeFazio:1999sv, Bosch:2004cb}. Here we present for the first time a simple
analytical expression for $S$. 

The shape function was first introduced to describe heavy quarks 
decaying while bound by quantum chromodynamics (QCD). It is employed in a
factorization formalism based on the heavy-quark effective field
theory (HQEFT) that separates the short-distance scale, related to the
heavy-quark mass, from the long-distance nonperturbative effects
governed by the scale $\Lambda_{\mathrm{QCD}}$, embodied in the shape
function. In QCD it
is a nonperturbative quantity that can  be fitted using data but
not yet derived theoretically. 

The shape function formalism has been defined also for quarkonium
\cite{Beneke:1997qw}. Subsequently in \cite{Beneke:1999gq} a
quarkonium production shape function was obtained
analytically.  Analytical results for the decay
shape function in the 't Hooft model were obtained in
\cite{Grinstein:2006pz}.

In QED, the shape function has recently been computed numerically and
applied to describe the decay of a muon bound in an atom
\cite{Czarnecki:2014cxa} (so-called decay in orbit, DIO). The spectrum
of decay electrons consists of the low-energy part up to about half
the muon mass $m_\mu$, and a (very suppressed) high-energy tail
extending almost to the full $m_\mu$. The shape function formalism
applies only to the former, also known as the Michel region
\cite{Michel:1949qe}.

In this paper we will not be concerned with the high-energy
tail. We note here only that it is also of great current interest
because it will soon be precisely measured  by  COMET \cite{Kuno:2013mha} and Mu2e
\cite{Brown:2012zzd}. The high-energy part of the DIO spectrum is a
potentially dangerous background for the exotic muon-electron conversion
search, the main goal of these experiments. That region has therefore
recently been theoretically scrutinized \cite{Czarnecki:2011mx, Szafron:2015kja}.

\section{Factorization in Muonium}
The HQEFT is based on the heavy quark mass being much
larger than the nonperturbative scale $\Lambda_{\mathrm{QCD}}$. Similarly, in 
muonic bound states there exists a hierarchy of scales \cite{Szafron:2013wja}:
the mass of the decaying muon is much larger than the typical residual momentum, $m_\mu
\gg p$. In a muonic atom we have 
\begin{equation}\label{eq:pa}
p \sim m_\mu Z\alpha,
\end{equation} 
while in muonium
\begin{equation}\label{eq:pm}
p \sim m_e \alpha,
\end{equation} where
$\alpha\approx {1}/{137}$
 is the fine structure constant and $m_e$
is the electron mass. 

With this observation, the factorization formula and the shape
function for the muon DIO were derived in \cite{Czarnecki:2014cxa}
using earlier QCD results \cite{Neubert:1993ch, Neubert:1993um,
  Mannel:1994pm, DeFazio:1999sv, Bosch:2004cb}.  Here we  follow
an equivalent but a slightly more general approach 
\cite{Bigi:1993ex} to derive the differential rate for a heavy charged
particle decay in the presence of an external Coulomb field,
neglecting radiative effects. We apply the result to find the decay
spectrum of muonium.

We  
concentrate on the  muon decay
$\mu^+ \rightarrow e^+ \bar{\nu}_\mu \nu_e$ but our results are
general and apply to any QED bound state decay,
provided that the momentum in the bound state is much
smaller than the decaying particle mass. 

\begin{figure}[ht]
\includegraphics[width=.7\columnwidth]{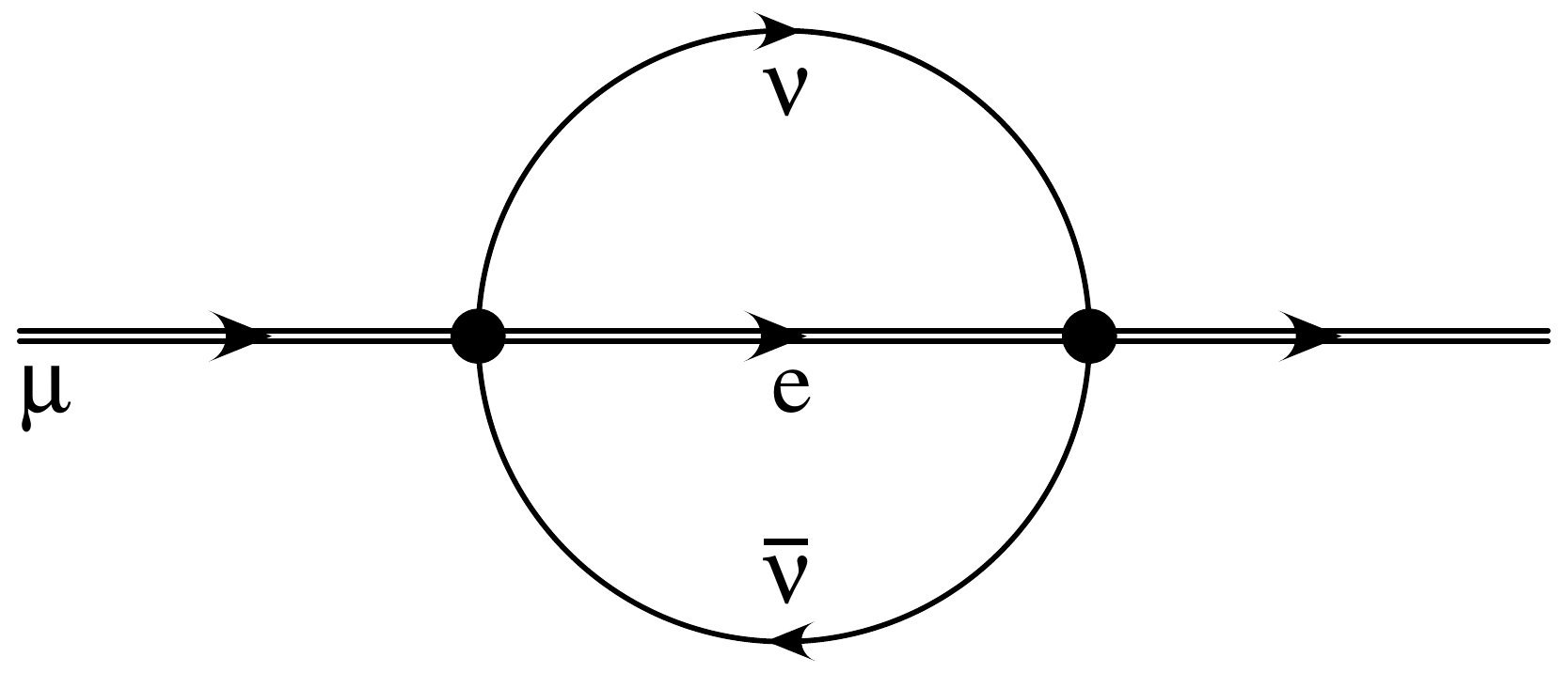}
\caption{\label{fig:dia1}
Muon self-energy diagram whose imaginary part corresponds to
  the muon decay rate. Double line for charged particles indicates the
  electromagnetic interaction with the spectator electron that needs to be resummed.}
\end{figure}

The decay amplitude is related
to the imaginary part of the two-loop diagram depicted in Fig.~\ref{fig:dia1}. 
Integrating over the relative neutrino momentum we express the
differential decay rate as
\begin{equation}\label{eq:spec}
\mathrm{d}\Gamma=2G_{F}^{2}\mathrm{Im}\left(h_{\alpha\beta}\right)W^{\alpha\beta}\frac{\mathrm{d}^{4}q}{(2\pi)^{3}},
\end{equation}
where $q$ is the sum of neutrino four-momenta and $G_F$ is the Fermi
constant  \cite{Webber:2010zf,Marciano:1999ih}.  The neutrino tensor is
\begin{equation}\label{eq:nute}
W^{\mu\nu}=-\frac{\pi}{3(2\pi)^{3}}q^{2}\left( g^{\mu\nu}-\frac{q^{\mu}q^{\nu}}{q^{2}}\right).
 \end{equation}
The charged-particle tensor $h_{\mu\nu}$ can be decomposed using
five scalar functions that depend on $q^2$ and $v\cdot q =q_0$. Here
$v$ is the four-velocity of the bound state.
In general,
\begin{eqnarray}
h_{\mu\nu}=-h_{1}g_{\mu\nu}+h_{2}v_{\mu}v_{\nu}-ih_{3}\epsilon_{\mu\nu\rho\sigma}v^{\rho}q^{\sigma}\nonumber\\+h_{4}q^{\mu}q^{\nu}+h_{5}\left(q_{\nu}v_{\mu}+q_{\mu}v_{\nu}\right),
 \end{eqnarray}
but since the neutrino tensor (\ref{eq:nute}) is symmetric under $\mu
\leftrightarrow \nu $, from now on we neglect the asymmetric part of $h$. 
Contracting the tensors and denoting $w_i=\mathrm{Im}(h_i)$ we find that only
two functions $w_{1,2}$ suffice to describe the double differential
spectrum,
\begin{equation}\label{eq:specdiff}
\frac{\mathrm{d}\Gamma}{\mathrm{d} q^2 \mathrm{d} q_0}=\frac{ G_{F}^{2}}{3\left(2\pi\right)^{4}}\left[3q^{2}w_{1}-\left(q^{2}-q_{0}^{2}\right)w_{2}\frac{}{}\right]\sqrt{q_0^2-q^2}.
\end{equation}

Functions $w_i$ can be calculated in QED.  Adopting  Schwinger's
operator representation \cite{Schwinger:1989ka},  we have instead of the free
electron propagator 
\begin{equation}
\frac{1}{\slashed k-m_e}\rightarrow \frac{1}{\slashed k +\slashed \pi -m_e},
\end{equation}
where $\pi^\mu$ is defined such that it does not contain
any heavy degrees of freedom. The commutator of its components gives 
the electromagnetic field-strength tensor $[\pi^\mu,
\pi^\nu]=-ieF^{\mu\nu}$ where $e$ is the muon charge. Formally,
\begin{equation}\label{eq:hmn}
h_{\mu\nu}=2\left\langle M \left|\overline{\mu}\gamma_{\mu}\frac{1}{\slashed
  k+\slashed\pi-m_e}\gamma_{\nu}\left(1-\gamma_{5}\right)\mu \right| M\right\rangle,
\end{equation}
where $|M\rangle$ denotes  the bound-muon state and $k=m_\mu v
-q$. Equation~(\ref{eq:hmn}) is
valid in the whole phase space.

To simplify our considerations, we restrict ourselves to the Michel
region where the electron is almost on-shell, $k^2\sim m_\mu \, p$ and
the time component of $k$ is large, $v\cdot k \gg p$. This is the
region where binding effects are most prominent. (Near the highest
energies also the virtuality is much higher, $k^2\sim m_\mu^2$,
permitting a perturbative expansion of the electron propagator
\cite{Szafron:2015kja}.) We  neglect the electron mass since the
electron is highly relativistic \cite{Mannel:1994pm,
  Shifman:1995dn}. The only effect of the electron mass is an overall
shift of the endpoint spectrum, just like in a free-muon decay.

In the Michel region, the four-momentum $k$ can be written as $k=(v\cdot
k)n+\delta k$, where $n$ is a lightlike vector,
$n^2=0$, and $\delta k \sim p$. Neglecting terms suppressed by $\frac{p^2}{m^2_\mu}$,
\begin{eqnarray}\label{eq:hmn_exp}
h_{\mu\nu}=4\left(2m_\mu v_{\mu}v_{\nu}-\nu\cdot
  kg_{\mu\nu}-v_{\nu}q_{\mu}-v_{\mu}q_{\nu} \right) \nonumber \\ \times
\left\langle M\left|\frac{1}{k^{2}+2\left(\pi\cdot n\right)\left(k\cdot v\right)}\right|M\right\rangle.
\end{eqnarray}
We cannot further expand the denominator since
both terms are of order $m_\mu\, p$.
We introduce $\lambda=-\frac{k^2}{2k\cdot v}$; it will  be useful
to remember that $\lambda$ scales like the muon momentum $\lambda \sim
p\sim Z\alpha$. We now define the shape function,
\begin{eqnarray}\label{eq:shape}
S(\lambda) = \left\langle M | \delta(\lambda-n \cdot \pi) | M \right\rangle,
\end{eqnarray}
and obtain
\begin{eqnarray}
w_{1}	&=&	2\pi S(\lambda),\\
w_{2}	&=&	\frac{4m_\mu}{k\cdot v}\pi S(\lambda)=\frac{2m_\mu}{k\cdot v}w_{1}.
\end{eqnarray}
We have recovered the QCD scaling behavior \cite{Bjorken:1968dy}:
functions $w_i$ depend in the leading order only on the ratio of
$k^2$ and $v\cdot k$ rather than on these two variables separately. 

Equation~(\ref{eq:shape}) reveals that the shape function is closely related to
the momentum distribution of the muon in the bound state. However, due
to  gauge invariance we cannot replace $n \cdot \pi$ by the  momentum
 in the $\vec{n}$ direction. 
\section{Shape function}
Formula (\ref{eq:shape}) is the
same for muonium and for a muonic atom. Both systems
 are nonrelativistic, therefore the
wave function needed to calculate the expectation value in (\ref{eq:shape}) has the same
analytical form. The only difference is its parameters and thus the
physical scales that characterize the muon momentum $p$ [see below, Eq.~(\ref{sub})].  We now
proceed to  an explicit calculation of the function $S$ in Eq.~(\ref{eq:shape}).

The bound-state wave function follows from  field theory via
the Bethe-Salpeter
equation \cite{Salpeter:1951sz}. In the non-relativistic
limit it reduces to the Schr\"{o}dinger
equation, 
\begin{equation}\label{eq:Sch}
\left(\frac{\vec{p}^{\,2}}{2\mu} +V(r)\right)\psi_S(r)=E \psi_S(r),
\end{equation}
where $\mu$ is the reduced mass of the system. In the
case of a muonic atom, with the mass of the nucleus $m_N$,
\begin{equation}\label{eq:mred}
\mu = \frac{m_\mu m_N}{m_\mu +m_N}\approx m_\mu.
\end{equation}
Subsequent formulas apply to muonium with the
following substitutions,
\begin{eqnarray}
m_\mu &\rightarrow& m_e,\nonumber \\ 
m_N &\rightarrow& m_\mu,\nonumber \\ 
Z &\rightarrow& 1.
\label{sub}
\end{eqnarray}
For example, the  reduced mass in muonium is
\begin{equation}\label{eq:mredm}
\mu = \frac{m_e m_\mu}{m_e +m_\mu}\approx m_e.
\end{equation}
With this notation we also have $p \sim Z\alpha\mu$.

As customary, Eq.~(\ref{eq:Sch})  is written in the Coulomb
gauge, with the electromagnetic four-potential given by 
\begin{equation}
eA_{\mu}(x)=\left(V(r),0,0,0\right),
\end{equation}
with $V(r) =-\frac{Z\alpha}{r}$ for a muonic atom or $V(r)
=-\frac{\alpha}{r}$ for muonium. 
The determination of the shape function  is especially convenient
in  the so-called light-cone gauge, 
\begin{equation}\label{eq:gaugecon}
n^\mu A_\mu(x)=0.
\end{equation}
In this gauge, the  electron is
effectively free up to effects quadratic in the electromagnetic field.
The price for this
simplification is a  more complicated formula for the muon wave function. 
In the light-cone gauge, Eq.~(\ref{eq:shape}) takes a simple form in the momentum representation,
\begin{equation}\label{eq:sx}
S(\lambda) = \int \frac{{\mathrm d}^3 k}{(2\pi)^3} \psi_S ^\star\left(\vec{k}\right)
\delta\left(\lambda+\vec{n}\cdot \vec{k}\right)\psi_S\left(\vec{k}\right).
\end{equation}
We are neglecting terms of order $(Z\alpha)^2$ in the above expression. 
To fulfill condition (\ref{eq:gaugecon}), we change the gauge,
\begin{equation}
eA_{\mu}'(x)=eA_{\mu}(x)+\partial_{\mu}\chi(x),
\end{equation}
with
\begin{equation}
\chi(x)=\chi(\vec{x})=Z\alpha\ln\left(\vec{n}\cdot \vec{r} +r\right).
\end{equation}
This transformation changes the muon Schr\"{o}dinger wave function
in the 1S state, $\psi_S(r)$, by an $\vec{r}$-dependent phase factor, such that
\begin{equation}\label{eq:psi}
\psi_{S}(r)\rightarrow\psi(\vec{r})=e^{-i\chi(\vec{r})}\psi_{S}(r)=\left(\vec{n}\cdot\vec{r}+r\right)^{-iZ\alpha}\psi_{S}(r).
\end{equation}
After the transformation, the wave function is no longer
 rotationally invariant, since the gauge fixing distinguishes
the 
direction of the outgoing electron. 

We use  the Schwinger parametrization,
\begin{equation}
\frac{\Gamma(\alpha)}{A^\alpha} =\int_0^\infty dt t^{\alpha -1}   \exp\left[-A t \right]
\end{equation}
to Fourier-transform Eq.~(\ref{eq:psi}),
\begin{widetext}
\begin{eqnarray}
\psi(\vec{k})&=& \int d^3 r\exp\left(-i\vec{k}\cdot \vec{r}\right)
\psi(\vec{r}) \nonumber\\ &=& \frac{iZ\alpha}{\Gamma(iZ\alpha)\sin \left(i\pi Z\alpha\right)}\frac{8\sqrt{\mu^3 Z^3
    \alpha^3
    \pi^3}}{\left(\mu^2Z^2\alpha^2+\vec{k}^2\right)^2}\left(\frac{\mu^2Z^2\alpha^2+\vec{k}^2}{2\left(\mu Z \alpha-i\vec{n}\cdot
  \vec{k}\right)}\right)^{iZ\alpha}
\left[ \frac{\mu^2 Z^2\alpha^2+\vec{k}^2}{2\left(\mu Z \alpha-i\vec{n}\cdot
  \vec{k}\right)}-\mu
                            (i+Z\alpha)\right].
\end{eqnarray}
We integrate in (\ref{eq:sx}) first 
over $\vec{n}\cdot \vec{k}$ using the delta-function,  then 
over $\vec{k}_\perp$, components of $\vec{k}$ perpendicular to $\vec{n}$,
\begin{equation}\label{eq:shape_res}
S(\lambda)=
\frac{2 \mu ^3 Z^6\alpha^6}{3\sinh(\pi  Z\alpha) 
}
\frac{
3 \lambda ^2+6 \lambda  \mu
   +\mu ^2 \left(4+Z^2\alpha^2\right) }{     \left[\lambda^2+\mu ^2 Z^2\alpha^2 \right]^3}
e^{2
   Z\alpha  \arctan\left(\frac {\lambda }{\mu  Z\alpha}\right)}.
\end{equation}
\end{widetext}
The exponential function in \eqref{eq:shape_res} arises from
$\left|\left(\mu Z \alpha + i\lambda\right)^{-i Z\alpha}\right|^2$,
appearing
after integrating $|\psi(\vec{k})|^2$ with the
delta-function in \eqref{eq:sx}.
The leading behavior can be  understood with the help of integral
\begin{equation}
\int d^2 k_\perp
\frac{1}{\left(\vec{k}^2_\perp+\lambda^2+\mu^2Z^2\alpha^2\right)^4}
\sim \frac{1}{(\lambda^2+\mu^2Z^2\alpha^2)^3}.
\end{equation}
The result (\ref{eq:shape_res}) contains subleading terms, related to the
 Coulomb potential in (\ref{eq:shape}), required
by the gauge invariance. At the current stage of calculations
$S(\lambda)$ is explicitly gauge independent. We can drop subleading
terms to obtain a leading-order formula,
\begin{equation}\label{eq:shape_res_exp}
S(\lambda)=
\frac{8 \mu ^5 Z^5\alpha^5 }{3\pi \left[\lambda^2+\mu ^2 Z^2\alpha^2 \right]^3}.
\end{equation}

The analytical formula obtained here is useful for several
reasons. First of all, counting powers and remembering that
$\lambda\sim p$, we find that $S(\lambda)\sim \frac{1}{p}\sim
\frac{1}{Z\alpha}$. This reminds us that $S(\lambda)$ is a
nonperturbative object and explains why its effect on the
spectrum can be quite dramatic, as we shall see in muonium in Sec.~\ref{sec:spec}.

Further, Eq.~(\ref{eq:shape_res}) allows us to better control the expansion and the
resummation of the $p$ effects
in the decay spectrum. This cannot be done so
easily with a numerical evaluation \cite{Czarnecki:2014cxa}, as is
especially clear when we analyze the first three moments, useful in  HQEFT for
constraining possible forms of the shape function.

The zeroth order moment gives just the  normalization. With the
 normalized wave function in Eq.~(\ref{eq:sx}), the shape
functions (\ref{eq:shape_res}) and (\ref{eq:shape_res_exp}) are
automatically normalized to unity; this is a consequence of the definition (\ref{eq:shape}).

When the subleading terms are neglected the first moment of the shape
function (\ref{eq:shape_res_exp}) vanishes,
\begin{equation}\label{eq:first}
\int \mathrm{d}\lambda\lambda S(\lambda) = \langle n\cdot \pi \rangle =0+\mathcal{O}(Z^2\alpha^2).
\end{equation}
Naive power counting in the left-hand side suggests a result
linear in $Z\alpha$. That leading part vanishes, similarly
to the  first moment  of the B-meson shape function. A contribution
linear in $\frac{p}{m_\mu}\sim Z\alpha$ is absent due
to the CGG/BUV theorem \cite{Chay:1990da, Bigi:1992su}. Moments of the
shape function are related to  matrix elements of local operators
in the heavy particle effective  theory.  Operators of dimensions 3
and 5 exist. A dimension 4 operator that
could generate, in the leading order, a nonvanishing first moment is
missing. A nonzero first moment can only appear at the
subleading order \cite{Neubert:1993um, Neubert:1993ch}.


The second moment is related to the square of the average momentum in the direction of $\vec{n}$,
 \begin{equation}\label{eq:second}
\int \mathrm{d}\lambda \lambda^2S(\lambda) =\frac{1}{3} \left\langle \vec{p}^{\,2} \right\rangle +\mathcal{O}\left(Z^4\alpha^4\right).
\end{equation}
In contrast to the first moment, there is no cancellation here and the
naive counting correctly predicts a result quadratic in $Z
\alpha$.
Therefore we do not need subleading corrections in
Eq.~(\ref{eq:hmn_exp}) to calculate (\ref{eq:second}).  This
moment characterizes the width $\sigma_\lambda$ of the
region smeared due to the shape function effects. As expected,
it is of the same  order as $p$:
$\sigma_\lambda = \frac{Z\alpha \mu}{\sqrt{3}}$.  This is similar to
the HQEFT where the second moment is also related to the average
kinetic energy of the heavy quark inside a meson.

In muonic aluminum, the stopping target of  the planned conversion
searches (Mu2e and
COMET), the shape function effect is sizeable
since $\sigma_\lambda \sim 6 \,\text{MeV}$, and has been precisely
measured by  TWIST  \cite{Grossheim:2009aa}. In the case of the muonium the
effect is much smaller, $\sigma_\lambda \sim 2 \, \text{keV}$, and is
negligible except near the end of the spectrum. 

\begin{figure}[ht]
    \centering
\vspace{4mm}
    \includegraphics[width=.99\columnwidth, trim = 0 0mm 0 0 ]{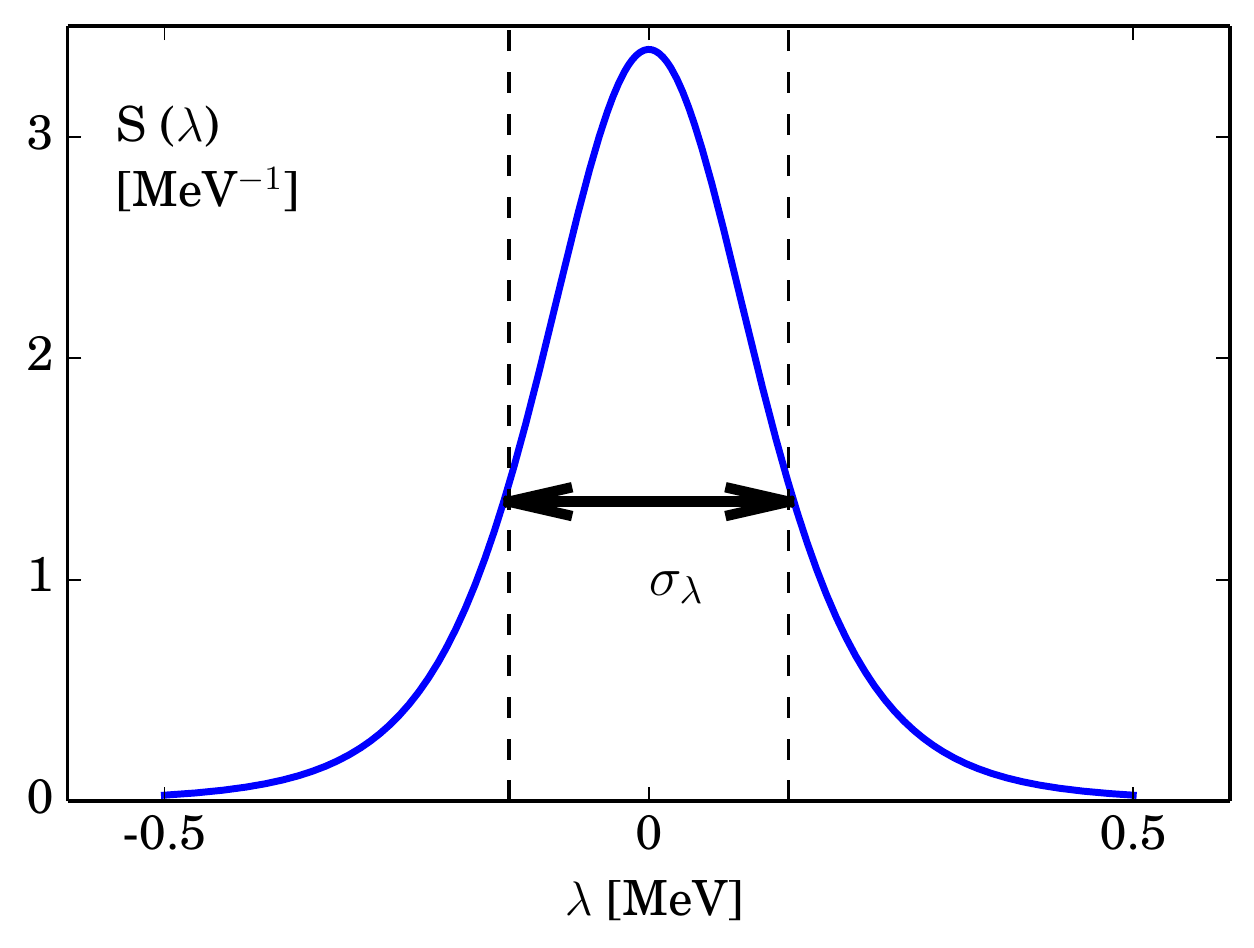}
    \caption{Leading shape (\ref{eq:shape_res_exp}) function for $\mu=1$ MeV and $Z\alpha
      =0.25$.  The width is $ \sigma_\lambda
      =\mu \frac{Z\alpha}{\sqrt{3}}$. To illustrate it better we included two vertical
      dashed lines at the ends of $\sigma_\lambda$ region.}
    \label{fig:shape}
\end{figure}

In Fig.~\ref{fig:shape} we plot the shape function for
$Z\alpha=0.25$. The width is proportional to $p$,
suggesting that the dominant effect is due to the
muon motion in the initial state.

Finally, we would like to point out that the formula
(\ref{eq:shape_res_exp}) can  guide phenomenological
models of the shape function in QCD. Some QCD bound states can
be described with the help of effective theories similar to
nonrelativistic QED \cite{Caswell:1985ui, Lepage:1987gg}.  For
example, Ref.~\cite{Jin:1997aj} 
postulated a similar functional
form of the shape function,
\begin{equation}
S(\lambda) = N\frac{\lambda(1-\lambda)}{(\lambda-b)^2+a^2}\theta(\lambda)\theta(1-\lambda),
\end{equation} 
with parameters $a,b$ to be fitted from data.  Our function has a
higher power of the denominator, therefore does not require an
artificial restriction of its support by $\theta$ functions, because
its tails are sufficiently suppressed. 

\section{Muonium spectrum}\label{sec:spec}
Having obtained the shape function, we can calculate the
muonium spectrum using (\ref{eq:specdiff}). After an integration over $q^2$,
the electron energy is given by
$E_e=m_\mu-q_0+\mathcal{O}(Z^2\alpha^2)$. 
The shape function
formalism can be interpreted as a replacement of the zero-width 
on-shell relation for the electron by a finite-width shape function
$S(\lambda)$. (If $S(\lambda)$ in the functions $w_i$ is replaced by
the  Dirac-delta on-shell condition, the free-muon decay spectrum results.)
Since the muon is almost at rest, the smearing is negligible far from the free muon
decay endpoint,  the only region where the
spectrum is  quickly varying  with the electron energy. 

We ignore the tail of the spectrum at energies higher than the free endpoint plus
several $\alpha m_e$. It is very suppressed and its evaluation
requires  perturbative corrections due to hard photons \cite{Szafron:2013wja,
  Szafron:2015kja}. We also ignore the lowest region of the spectrum
where positronium can be formed \cite{Greub:1994fp}.

\begin{figure}[htb]
    \centering
\vspace{4mm}
    \includegraphics[width=.99\columnwidth, trim = 0 0mm 0 0]{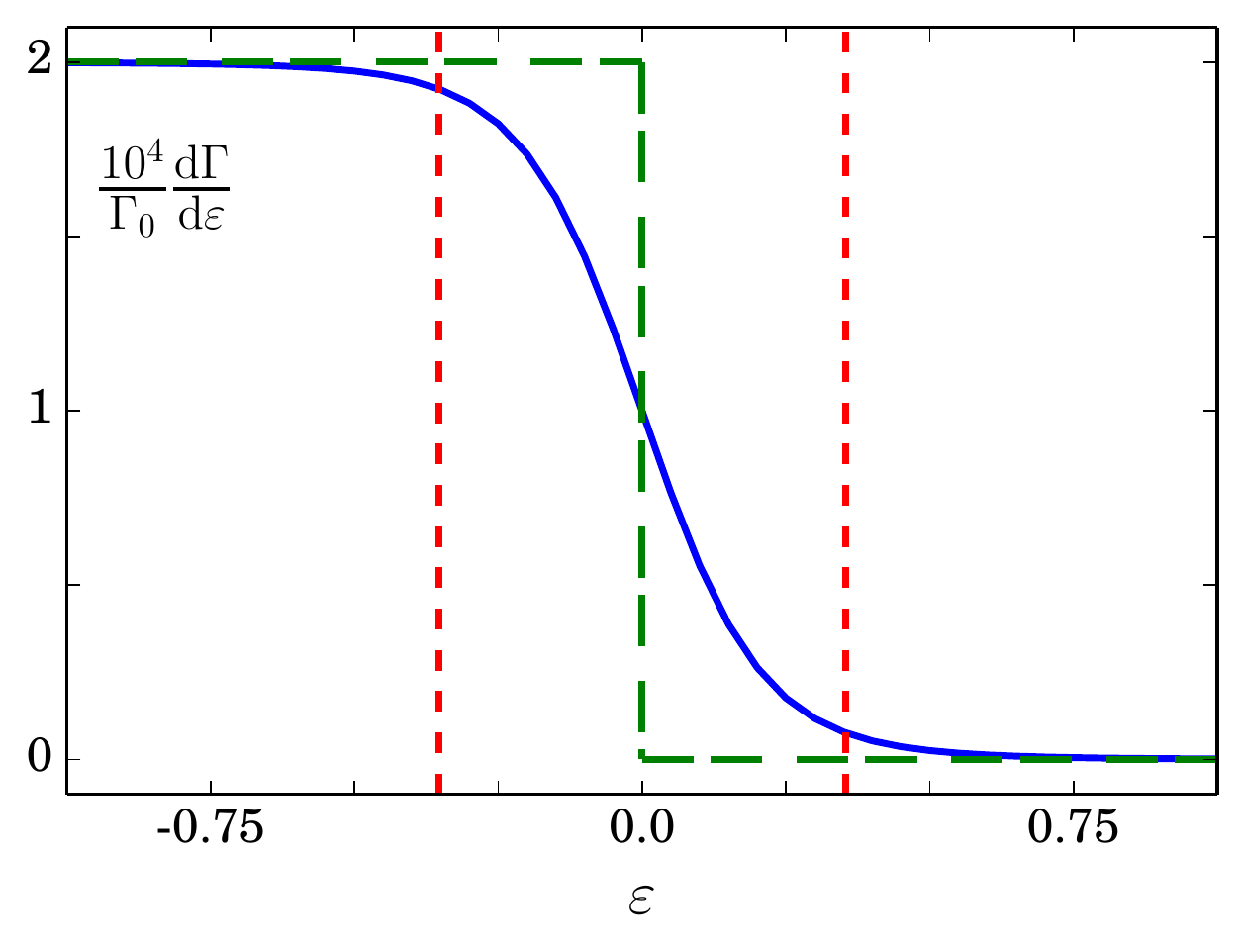}
    \caption{Endpoint region of the muonium decay. Electron energy is
      parametrized in terms of $\varepsilon$, such that $\varepsilon =
      10^4\frac{E_e-E_{\text{max}}}{E_{\text{max}}}$, $E_{\text{max}}=\frac{m_\mu^2+m_e^2}{2m_\mu}$. Dashed (solid) line shows the
      free-muon (muonium)
      spectrum.  Vertical dotted lines emphasize the size of
   the region that is smeared due to binding effects. }
    \label{fig:spectrum}
\end{figure}

For illustration, Fig.~\ref{fig:spectrum} shows the  muonium decay spectrum in the vicinity of the  free
muon endpoint. 
The extent of the region affected by the shape function corresponds
to the smearing width $\sigma_\lambda$, denoted by two vertical lines.
In this region the slope of the spectrum is proportional to the shape function $S(\lambda)$ and
therefore is of the order of $\frac{1}{p}\sim
\frac{1}{\sigma_\lambda}$. 

 Note that the free-muon decay, denoted
with the dashed line, resembles a step function. This is an artefact
of the very narrow width of the region shown in this figure. In fact,
the free-decay spectrum varies with $\varepsilon$ to the left of the
step and vanishes to the right of it.

\section{Conclusions}
We have derived an analytical formula for the shape function and
used it to calculate the   muonium decay spectrum. Shape
function moments were analyzed and compared with appropriate
expressions in HQEFT. Our analytical formula may also have a limited
application to  describe nonrelativistic QCD systems.
For now, the analytical expression for the shape function has improved our
understanding of 
the approximations used in the derivation of the muon DIO spectrum.

\begin{acknowledgments}
This research was supported by Natural Sciences and Engineering
Research Council (NSERC) of Canada.
\end{acknowledgments}


%

\end{document}